# A Theoretical Closure for Turbulent Flows Near Walls

Trinh, Khanh Tuoc

Institute of Food Nutrition and Human Health

Massey University, New Zealand

*K.T.Trinh@massey.ac.nz*

## Abstract

This paper proposes a simple new closure principle for turbulent shear flows. The turbulent flow field is divided into an outer and an inner region. The inner region is made up of a log-law region and a wall layer. The wall layer is viewed in terms of the well known inrush-sweep-burst sequence observed since 1967. It is modelled as a transient laminar sub-boundary layer, which obeys Stokes' solution for an impulsively started flat plate. The wall layer may also be modelled with a steady state solution by adding a damping function to the log-law.

Closure is achieved by matching the unsteady and steady state solutions at the edge of the wall layer. This procedure in effect feeds information about the transient coherent structures back into the time-averaged solution and determines theoretically the numerical coefficient of the logarithmic law of the wall

The method gives a new technique for writing accurate wall functions, valid for all Reynolds numbers, in computer fluid dynamics (CFD) programmes.

## 1    Introduction

Turbulence is a complex time dependent three-dimensional motion widely believed to

be governed by equations[1] established independently by Navier and Stokes more than 150 years ago

$$\frac{\partial}{\partial t}(\rho u_i) = -\frac{\partial}{\partial x_i}(\rho u_i u_j) - \frac{\partial}{\partial x_i}p - \frac{\partial}{\partial x_i}\tau_{ij} + \rho g_i \tag{1}$$

and the equation of continuity

$$\frac{\partial}{\partial x_i}(\rho u_i) = 0 \tag{2}$$

The omnipresence of turbulence in many areas of interest such as aerodynamics, meteorology and process engineering, to name only a few, has nonetheless led to a voluminous literature.

Most of the interest in turbulence modelling from a practical engineering view point was originally based on the time averaged parameters of the steady state flow field. Reynolds (1895) has proposed that the instantaneous velocity $u_i$ at any point may be decomposed into a long-time average value $U_i$ and a fluctuating term $U'_i$.

$$u_i = U_i + U'_i \tag{3}$$

with

$$U_i = \lim_{t \to \infty} \int_0^t u_i \, dt \tag{3.1}$$

$$\int_0^\infty U'_i \, dt = 0 \tag{3.2}$$

For simplicity, we will consider the case when
    1. The pressure gradient and the body forces can be neglected
    2. The fluid is incompressible (ρ is constant).

Substituting equation (3) into (1) and taking account of the continuity equation (2) gives:

$$U_i \frac{\partial U_j}{\partial x_j} = v \frac{\partial^2 U_i}{\partial x_i} - \frac{\partial \overline{U'_i U'_j}}{\partial x_i} \tag{4}$$

---

[1] The suffices i and j in this paper refer to standard vector notation.

These are the famous Reynolds equations (Schlichting, (1960), p. 529) also called Reynolds-Averaged-Navier-Stokes equations RANS (Gatski and Rumsey, 2002, Hanjalić and Jakirlić, 2002). The long-time-averaged products $\overline{U'_i U'_j}$ arise from the non-linearity of the Navier-Stokes equations. They have the dimensions of stress and are known as the Reynolds stresses. They are absent in steady laminar flow and form the distinguishing features of turbulence.

The NS equations and the equation of continuity form a closed set that can be solved in principle, even though no general solution has been obtained in the last 160 years because of the great difficulties arising from the non linear terms. When Reynolds averaged the NS equations a degree of freedom is lost and there is no longer sufficient information to solve this new set of equations. This is the famous closure problem in turbulence. It is solved currently by formulating the Reynolds stresses with empirical or semi theoretical models. There are many ways to address this closure problem (Hinze, 1959, Schlichting, 1960, McComb, 1991, Lesieur, 2008).

One the first attempts was made by Prandtl (1925, 1935) which resulted in his formulation of the logarithmic law of the wall. The critical step in Prandtl's analysis is the derivation of a turbulent length scale $l$ called 'mixing-length' that allowed him then to estimate the eddy viscosity. The mathematical success of his approach, verified with the data of Nikuradse (1932) was marred however by his unfortunate analogy of the scale $l$ with the free path of molecules. While the physical interpretation of this scale is now widely discredited the law of the wall remains one of the most useful tools in turbulence predictions. The postulation of an algebraic relationship for the length scale ready for use with the RANS is referred to in the literature as a zero-order closure model (Gatski and Rumsey, 2002). Prandtl's mixing length was postulated for the plane shear flow with unidirectional mean flow. In addition, Prandtl focused only on the log-law region and made no attempt to model the outer region or the wall layer. Prandtl did assume that there was a thin region near the wall $0 < y^+ < 5$, called the laminar sublayer where viscous forces would completely dominate resulting, in his view, to complete damping of turbulent fluctuations. Many other workers have extended the zero order analysis to two-layer mixing-length models. Cecebi & Smith (1974) essentially expressed Prandtl's law and the law of the wake (Coles, 1956) in multidimensional terms and Baldwin and Lomax (1978) expressed the mixing length in terms of vorticity.

The considerable difficulties linked with analytical solutions for flow in complex geometries were by-passed with the introduction of computational fluid dynamics CFD in the early seventies (Launder and Spalding, 1974). The problem with zero order models is that the parameters of the model e.g. the boundary layer thickness in the Cecebi-Smith model must be evaluated by searching along grid lines in the normal direction. One equation models such as that of Spalart & Allmaras (1994) are local and can be used with any type of grid. The approach here is to calculate the eddy viscosity through the formulation of a transport equation. The next development was to calculate the eddy viscosity from two local quantities both estimated from transport equations.

$$E_v = C_\mu \frac{k^2}{\varepsilon} \tag{5}$$

Where the symbols here are

$E_v$  Turbulent kinematic viscosity

$k$  Turbulent kinetic energy

$\varepsilon$  Turbulent energy dissipation rate

$C_\mu$  Model coefficient

This is the famous and popular $k-\varepsilon$ model. An alternative is the $k-\omega$ model where $\omega$ is the dissipation rate of kinetic energy per unit $k$. The coefficients of the terms in the transport equations for these models are determined empirically from experimental observations. One well-known challenge of the $k-\varepsilon$ models is the treatment of the inner region. Bradshaw, Launder, & Lumley (1991) tested CFD packages developed by authors around the world and found that any model which invokes the logarithmic law-of-the-wall gave reasonable predictions of the velocity field irrespective of the model for the outer flow. But the so-called standard $k-\varepsilon$ models, those derived for high Reynolds numbers, do not give an accurate representation of recirculation regions and of the near wall at transition and low Reynolds numbers. Thus there is an extensive list of modifications of Prandtl's law of the wall to deal with these situations e.g. (Gavrilakis, 1992) which has also been shown to apply to two dimensional flow (Zanoun and Durst, 2003). A further weakness of the $k-\varepsilon$ models is the unrealistic isotropic nature of the eddy viscosity. This has led to the development of so-called non-linear eddy viscosity models such as the algebraic stress models.

The latest development is the introduction of differential second-moment turbulence closure models DSM (Launder and Sandham, 2002) that are based on transport equations for the turbulent stresses and turbulent fluxes. The advantage of the DSM is in the exact treatment of the turbulence production term and of anisotropy of the turbulent stress field. Hanjalíc and Jakirlić (2002) believe that the DSM will eventually replace the present popular $k-\varepsilon$ model but admit that "despite more than three decades of development and significant progress, these models are still viewed… as a …target than as a proven and mature technique". The modelling of the $\overline{U_i U_j}$ and $\varepsilon$ equations is still based on the a characteristic turbulence time scale $\tau = k/\varepsilon$ and a length scale $L = k^{3/2}/\varepsilon$ but one now has the opportunity to model two new important terms: the pressure-strain term $\varphi_{ij}$ and the stress dissipation rate

This very succinct overview of closure techniques is designed only to give context. There is no attempt to capture adequately, even in the most general manner, the huge diversity of approaches resulting from the avalanche of computer modelling work of the last forty years. For a more detailed introduction, the reader may consult excellent books and reviews such as that of Launder and Sandham (op. cit.) or (Patel et al., 1985, Rodi, 1980) for earlier models.

The striking feature of all existing closure models is the empirical nature of coefficients used, which are more and more numerous as the models increase in complexity to give adequate descriptions of complex industrial applications. There were of course considerations of basic theoretical understanding, particularly of the energy cascade introduced by Kolmogorov, but the fundamental empirical nature simply reflects the state of poor understanding of turbulence mechanisms and, in my view, the constraints of the RANS used as a starting point.

This paper proposes a new method for effecting closure in turbulence modelling.

## 2     Physical Visualisation And Principle Of The Closure Technique

Turbulent flow fields near walls can be divided into an outer and an inner region e.g. (Panton, 1990). The inner region can be further divided into a wall layer and a log-law layer following the work of Prandtl and Millikan.

### 2.1    The wall layer

The process of turbulence production in the wall layer has been well documented (Kline et al., 1967, Kim et al., 1971, Offen and Kline, 1974, Corino and Brodkey, 1969). Despite the dominating influence of viscous momentum, the flow field near the wall is not laminar in the steady-state sense, but highly active. Periodically, fast fluid rushes from the outer region towards the wall then follows a vortical sweep along the wall The travelling vortex induces underneath its path (Figure 1) which is observed as streaks of low-speed fluid. The streaks tend to lift, oscillate and eventually burst in violent ejections from the wall towards the outer region. The low speed streak phase is much more persistent than the ejection phase and dominates the contribution to the time-averaged profile (Walker et al., 1989). The wall layer process has been modelled in so called kernel studies that investigate the interactions between a moving vortex and the wall below its path (Walker, 1978, Hall and Horseman, 1991, Peridier et al., 1991, Swearingen and Blackwelder, 1987, Harvey and Perry, 1971, Ersoy and Walker, 1986, Chu and Falco, 1988, Liu et al., 1991, Tucker and Conlisk, 1992). An alternative approach is to treat the low-speed streaks in the sweep phase as a simple two dimensional sinusoidal flow sometimes called Kolmogorov flow (Obukhov, 1983), or better still analyse them with techniques borrowed from laminar oscillating flow (Trinh, 2009b). For this purpose, the local instantaneous velocity is written as

$$u_i = \widetilde{u}_i + u'_i \qquad (6)$$

Where $\widetilde{u}_i$ smoothed velocity which evolves within the period $t_v$ of the wall layer (sweep phase) and $u'_i$ fast fluctuations of period $t_f$. Introducing the time-averaged velocity

$$u_i = U_i + \widetilde{U}'_i + u'_i \qquad (7)$$

where

$$\widetilde{U}'_i = \widetilde{u}_i - U_i \tag{8}$$

We may average the Navier-Stokes equations over the period $t_f$ of the fast fluctuations. Bird, Stewart, & Lightfoot (1960), p. 158 give the results as

$$\frac{\partial(\rho \widetilde{u}_i)}{\partial t} = -\frac{\partial p}{\partial x_i} + \mu \frac{\partial^2 \widetilde{u}_i}{\partial x_j^2} - \frac{\partial \widetilde{u}_i \widetilde{u}_j}{\partial x_j} - \frac{\partial \overline{u'_i u'_j}}{\partial x_j} \tag{9}$$

Equation (9) defines a second set of Reynolds stresses $\overline{u'_i u'_j}$ which we will call "fast" Reynolds stresses to differentiate them from the standard Reynolds stresses $\overline{U'_i U'_j}$. We may write the fast fluctuations in the form

$$u'_i = u_{0,i}\left(e^{i\omega t} + e^{-i\omega t}\right) \tag{10}$$

The fast Reynolds stresses $u'_i u'_j$ become

$$u'_i u'_j = u_{0,i} u_{0,j} (e^{2i\omega t} + e^{-2i\omega t}) + 2 u_{0,i} u_{0,j} \tag{11}$$

Equation (11) shows that the fluctuating periodic motion $u'_i$ generates two components of the "fast" Reynolds stresses: one is oscillating and cancels out upon long-time-averaging, the other, $u_{0,i} u_{0,j}$ is persistent in the sense that it does not depend on the period $t_f$. The term $u_{0,i} u_{0,j}$ indicates the startling possibility that a purely oscillating motion can generate a steady motion which is not aligned in the direction of the oscillations. The qualification steady must be understood as independent of the frequency $\omega$ of the fast fluctuations. If the flow is averaged over a longer time than the period $t_v$ of the bursting process, the term $u_{0,i} u_{0,j}$ must be understood as transient but non-oscillating. This term indicates the presence of transient shear layers embedded in turbulent flow fields and not aligned in the stream wise direction similar to those associated with the streaming flow in oscillating laminar boundary layers (Schneck and Walburn, 1976, Tetlionis, 1981).

Thus the local instantaneous velocity needs to include an extra term to account for this streaming flow that we missed in equation (6)

$$u_i = \widetilde{u}_i + u'_i + u_{i,st} \tag{12}$$

or

$$u_i = U_i + \widetilde{U}'_i + u'_i + u_{i,st} \tag{13}$$

The streaming flow represents the viscid-inviscid interaction in the kernel studies (e.g. Peridier et al. (op.cit.) or the ejections associated with the bursting phase. In their DNS of turbulent flow near a wall Schoppa and Hussain (2002) have similarly concluded that strong shear layers that they call transient stress growth TSG are produced by sinusoidal velocity fluctuations in the flow .

The velocity components in equation (13) can be obtained in principle by solving the NS equations expressed in terms of this four component instantaneous velocity (Trinh, 2009a) but the solution is extremely complex. None exists yet. We can progress can adopting the method of successive approximations used in the study of laminar oscillating flow (Schlichting, Tetlionis op.cit.). The smoothed phase velocity $\tilde{u}_i$ may be obtained by neglecting the effect of the fast fluctuations and therefore the terms $u'_i$ and $u_{i,st}$. This requires that the term

$$\varepsilon = \frac{U_e}{L\omega} \ll 1 \tag{14}$$

where $U_e$ is the local mainstream velocity and L is a characteristic dimension of the body. For flow past a flat plate the governing of this solution of order $\varepsilon^0$ is

$$\frac{\partial \tilde{u}}{\partial t} + \tilde{u}\frac{\partial \tilde{u}}{\partial x} + v\frac{\partial \tilde{v}}{\partial y} = v\frac{\partial^2 \tilde{u}}{\partial y^2} \tag{15}$$

where $v$ is the kinematic fluid viscosity and $\tilde{u}$ $\tilde{v}$ are the velocity components $\tilde{u}_i$ in the x and y directions. Equation (15) has further been simplified by Stokes (1851) for an impulsively started flow

$$\frac{\partial \tilde{u}}{\partial t} = v\frac{\partial^2 \tilde{u}}{\partial y^2} \tag{16}$$

Stokes has solved this equation for the conditions:
| | | | |
|---|---|---|---|
| IC | t = 0 | all y | $\tilde{u} = U_v$ |
| BC1 | t > 0 | y = 0 | u = 0 |
| BC2 | t > 0 | y = ∞ | $\tilde{u} = U_v$ |

where $U_v$ is the approach velocity for this sub-boundary layer. The velocity at any time t after the start of a period is given by:

$$\frac{\tilde{u}}{U_v} = \mathrm{erf}(\eta_s) \tag{17}$$

where $\eta_s = \dfrac{y}{\sqrt{4\nu t}}$

The average wall-shear stress is

$$\tau_w = \dfrac{\mu U_\nu}{t_\nu} \int_0^{t_\nu} \left(\dfrac{\partial \tilde{u}}{\partial y}\right)_{y=0} dt = \dfrac{\mu U_\nu}{t_\nu \sqrt{\pi}} \int_0^{t_\nu} \dfrac{1}{\sqrt{\nu t}} dt \qquad (18)$$

Equation (18) may be rearranged as

$$t_\nu^+ = \dfrac{2}{\sqrt{\pi}} U_\nu^+ \qquad (19)$$

The time-averaged velocity profile near the wall may be obtained by rearranging equation (17) as

$$\dfrac{U^+}{U_\nu^+} = \int \mathrm{erf}\left(\dfrac{y^+}{4 U_\nu^+ \sqrt{t/t_\nu}}\right) d\left(\dfrac{t}{t_\nu}\right) \qquad (20)$$

Equation (20) applies up to the edge of the wall layer where $u/U_\nu = 0.99$, which corresponds to $y = \delta_\nu$ and $\eta_s = 1.87$. Substituting these values into equation (17) gives

$$\delta_\nu^+ = 4.16 U_\nu^+ \qquad (21)$$

Back-substitution of equation (16) into (14) gives

$$\delta_\nu^+ = 3.78 t_\nu^+ \qquad (22)$$

where the velocity, period and normal distance have been normalised with the wall parameters $\nu$ the kinematic viscosity and $u_* = \sqrt{\tau_w/\rho}$ the friction velocity, $\tau_w$ the time averaged wall shear stress and $\rho$ the density.

The Stokes solution has been used by many authors to model the wall layer of turbulent flows (Einstein and Li, 1956, Hanratty, 1956, Black, 1969) and successfully predicted the time averaged velocity profile and the time scale $t_\nu$ of the wall layer. It can also reproduce successfully many of the classical turbulence statistics of turbulence: the probability density function, the fluctuating velocities, the production of turbulence, the correlation function and the energy spectrum (Trinh, 2010c) but a fundamental difference in the physical visualisation must be pointed out between the present approach and that of previous users of the Stokes solution. Equation (15) has been used by previous authors in an Eulerian framework. The low speed streaks are developing viscous layers that occur randomly in time and space. Thus a Lagrangian framework is better suited. Secondly, the time scale of the wall layer measured by Meek and Baer

(op.cit.) is far too large to assume that the flow is impulsively started. In fact, the convection terms in equation (14) are of the same order of magnitude as the velocity derivative (Trinh, 2010b) and there is no justification for neglecting them in an Eulerian framework. By rewriting equation (15) as

$$\frac{D\widetilde{u}}{Dt} = \nu \frac{D^2\widetilde{u}}{Dy^2} \qquad (23)$$

Where $D/Dt$ is a partial derivative along the path of diffusion of viscous momentum, we can decouple the effect of diffusion and convection (Trinh, 2010b) and thus apply Stokes solution formally. Thus equation (21) can be used to define the wall layer thickness as the position of maximum penetration of wall retardation by diffusion of viscous momentum.

The solution of order $\varepsilon^0$ is independent of the solutions of order $\varepsilon$ that deal mainly with the ejections and therefore the log-law and outer region. Jimenez and Pinelli (1999) have provided evidential support for that observation in their paper entitled The autonomous cycle of near-wall turbulence" where they show by DNS that events in the wall layer are not greatly affected by the structures in the log-law and outer regions.

## 2.2 The log-law layer and the outer region

As the vortex travels along the wall, the fluctuations imposed on the low speed streaks grow and the streaming flow contains substantial amount of kinetic energy sufficient to eject wall fluid from the wall layer. The ejections start to disturb the outer quasi-inviscid region beyond the wall layer and dramatically increase the boundary layer thickness. The magnitude of the parameter $\varepsilon$ is no longer negligible and we need to switch to a second approximation of order $\varepsilon$ to evaluate the velocities $u'_i$ and $u_{i,st}$. Unfortunately, the analytical solution of this subset of the NS equations is quite difficult and only numerical solutions are possible. These may give greater insight into turbulence mechanisms but are not helpful in the predictions of transport rates. Thus the most realistic approach is still to use the RANS for the outer region.

In simple flow geometries, the outer region has been modelled with phenomenological models such as the defect law of Karman (1934) and the law of the wake (Coles, 1956). In complex geometries the modelling is much better achieved by the use of computational fluid dynamics CFD packages. As Bradshaw et al. (op. cit.) the empirical model used for the outer region is not important as long as interfaces with the log-law near the wall. At the moment this interface varies between software designs very arbitrarily.

**2.3    Principle of the closure technique**

The critical point noted in this theory is that turbulence is a local instantaneous phenomenon and cannot be adequately explained by measurements and analysis of time-averaged shear stresses and velocities only. A great confusion has resulted from starting the theoretical analysis with time-averaged versions of the Navier-Stokes equations, the so-called Reynolds equations, whereas the present theory starts with the unsteady state Navier-Stokes equations and then time averages the solution. Of course, an unsteady state solution of the Navier-Stokes equation for the entire turbulent flow field is still out of reach. The problem is circumvented by taking a zonal approach. By matching the unsteady state solution for the wall layer and time-averaged solutions at the pivotal edge of the wall layer, closure of the time-averaged model is achieved formally.

**3    Theory**

The outer region scales with the outer variables $\delta$ the boundary layer thickness and $U_\infty$ the approach velocity but the wall layer scales with the wall variables $u_*$ and $\nu$. Millikan (1939) showed from similarity considerations that there must exist an intermediary layer where

$$U^+ = A \ln y^+ + B \tag{24}$$

Thus confirming the result of Prandtl (op.cit.) without the need for cumbersome postulates about the physical process.

We now attempt to derive a formulation compatible with Millikan's analysis, using a zero equation model to highlight the basic considerations without dealing with the

added complexities of higher models. The local time-averaged shear stress at a distance y is given by

$$\tau = (\mu + \rho E_v)\frac{dU}{dy} \tag{25}$$

where $E_v$ is called the eddy viscosity after a proposal by Boussinesq (1877). Rearranging in dimensionless form gives

$$U^+ = \int_0^{y^+} \frac{\tau/\tau_w}{1+E_v/\nu} dy^+ \tag{26}$$

The velocity changes over a distance l are correlated because the ejections are a coherent fluid structures that travels significant distances (Johansson et al., 1991) and the velocity may be expanded as a Taylor series in the positive and negative directions.

$$|\Delta U| = |U_1 - U| = |U_2 - U| = \left| l\frac{dU}{dy} \pm l^2 \frac{d^2U}{dy^2} + ... \right| \tag{27}$$

Neglecting second order and higher terms, Prandtl estimated the fluctuating velocities as

$$U' \approx V' \approx \Delta U \approx l\frac{dU}{dy} \tag{28}$$

and the turbulent Reynolds stresses and the eddy viscosity as

$$\tau_t = \overline{U'V'} = \rho l^2 \left(\frac{dU}{dy}\right)^2 = \rho E_v \frac{dU}{dy} \tag{29}$$

Hence

$$E_v = l^2 \frac{dU}{dy} = l\sqrt{\frac{\tau_t}{\rho}} \tag{30}$$

Equation (26) may be rearranged as

$$\frac{dU^+}{dy^+} = \frac{\tau/\tau_w}{1+E_v/\nu} \tag{31}$$

In the log law region, the viscous contribution is small and $1 << E_v/\nu$ and we may write

$$\frac{dU^+}{dy^+} \cong \frac{\tau/\tau_w}{E_v/\nu} = \frac{\tau/\tau_w}{(l/\nu)\sqrt{\tau_t/\rho}} = \frac{\tau/\tau_w}{(lu_*/\nu)\sqrt{\tau_t/\tau}\sqrt{\tau/\tau_w}} \tag{32}$$

Prandtl assumed that the scale l is proportional to the normal distance y

$$l = \kappa y \tag{33}$$

where $\kappa$ is known as Karman's constant. This constant may be interpreted as the inclination of the ejection with respect to the normal distance (Trinh, 1996). For the moment we simply accept the widely quoted value $\kappa = 0.4$ to avoid complicating the formulation any further. Substituting equation (33) into (32)

$$\frac{dU^+}{dy^+} = \frac{\sqrt{\tau/(\tau_w \tau_t)}}{\kappa y^+} \tag{34}$$

which does not lead to a log-law. To obtain the log law from equation (34) Prandtl had to assume further that

$$\tau = \tau_w = \tau_t \tag{35}$$

This assumption is clearly at odds with reality particularly at low Reynolds number and we therefore will not adopt it. In addition, we do not interpret $l$ as a mixing-length, a distance that eddies at y travel before they loose their identity in analogy to the mean free path of gas molecules but as a typical scale of the streaming jet. The lower limit of application of the log-law is the edge of the wall layer (Trinh, 2010a) In order to extend produce a single equation that applies both to the log-law and wall layers, we introduce a new factor F such that

$$l = \kappa y \sqrt{\frac{\tau}{\tau_w}} F \tag{36}$$

And

$$\frac{E_v}{\nu} = \kappa y^+ \left(\frac{\tau}{\tau_w}\right) \sqrt{\frac{\tau}{\tau}} F \tag{37}$$

Van driest (1956) proposed a similar a "damping factor" inspired from the Stokes solution for an oscillating flat plate. He argued that this solution can be used to describe how eddies from the outer region would be damped by viscous resistance near the wall. In the present approach the inclusion of a damping factor F is only to make equation (37) compatible with the solution of order $\varepsilon^0$ and give a single formulation for the inner region. It has no physical significance, and certainly I do not believe that the function F represents a damping of turbulent eddies from the outer flow that bombard the wall layer, a view that was made obsolete by the ground breaking work of (Kline, et al., 1967).

The form of this factor may be inferred from the velocity gradient in the Stokes

solution

$$\frac{\partial \tilde{u}}{\partial \tilde{y}} = \frac{2}{\sqrt{\pi}} e^{-\eta_s^2} \tag{38}$$

A form of the damping function compatible with the error function is

$$F = 1 - e^{-b\left[\frac{y^+(\tau/\tau_w)}{\delta_v^+}\right]^2} \tag{39}$$

The damping function must satisfy two limits

$F = 0$      at      $y^+ = 0$

$F \to 1$      as      $y^+ \to \delta_v^+$

As with all boundary layer solutions, an arbitrary cut-off value must be taken. In the present formulation a value F=0.99999 at $y^+ = \delta_v^+$ was found most suitable. At high Reynolds numbers, the wall layer is thin and $\delta_v^+ \ll R^+$ and $\tau/\tau_w \cong 1$, equation (39) gives $b = 11.2$. Then for pipe flow

$$U^+ = \int_0^{y^+} \frac{(1 - y^+/R^+) dy^+}{1 + 0.4 y^+ \left(1 - \frac{y^+}{R^+}\right) \sqrt{\frac{E_v}{E_v + v}} \left(1 - e^{-11.2\left[\frac{y^+(1-y^+/R^+)}{\delta_v^+}\right]^2}\right)} \tag{40}$$

Equation (40) must pass by the point $(\delta_v^+, U_v^+)$. Matching equations (40) and (21) gives

$$\delta_v^+ = 4.16 \int_0^{\delta_v^+} \frac{(1 - y^+/R^+) dy^+}{1 + 0.4 y^+ \left(1 - \frac{y^+}{R^+}\right) \sqrt{\frac{E_v}{1 + E_v}} \left(1 - e^{-11.2\left[\frac{y^+(1-y^+/R^+)}{\delta_v^+}\right]^2}\right)} \tag{41}$$

## 4     Comparison with literature data

This equation can be solved iteratively for $\delta_{v+}$ for each value of $R^+$. The effect of the Reynolds number is introduced through the dimensionless radius $R^+ = \frac{\text{Re}\sqrt{f}}{2\sqrt{2}}$. The wall layer thicknesses predicted by equation (41) are shown against experimental data in Figure1.

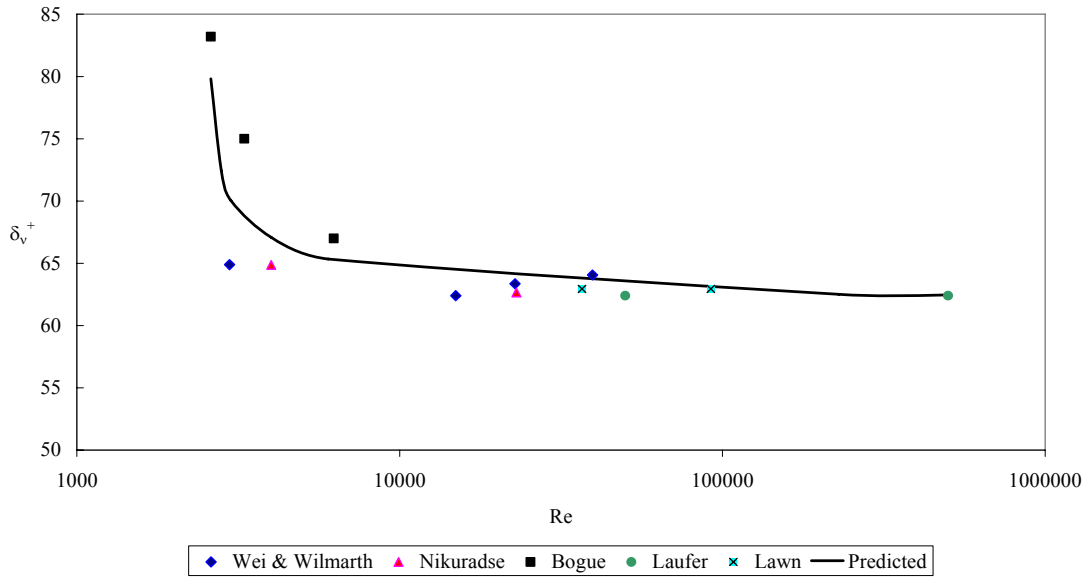

Figure 1 Comparison of wall layer thickness predicted by equation (41) with data from Wei and Wilmarth (1989), Nikuradse (Nikuradse, 1932), Bogue (Bogue, 1962), Laufer (1954) and Lawn (1971).

The velocity profiles predicted from equation (40) are plotted against literature data in Figure 2. The predictions fit the data in the inner region (wall layer + log law) but not in the pipe core which requires a different correlation than the log-law.

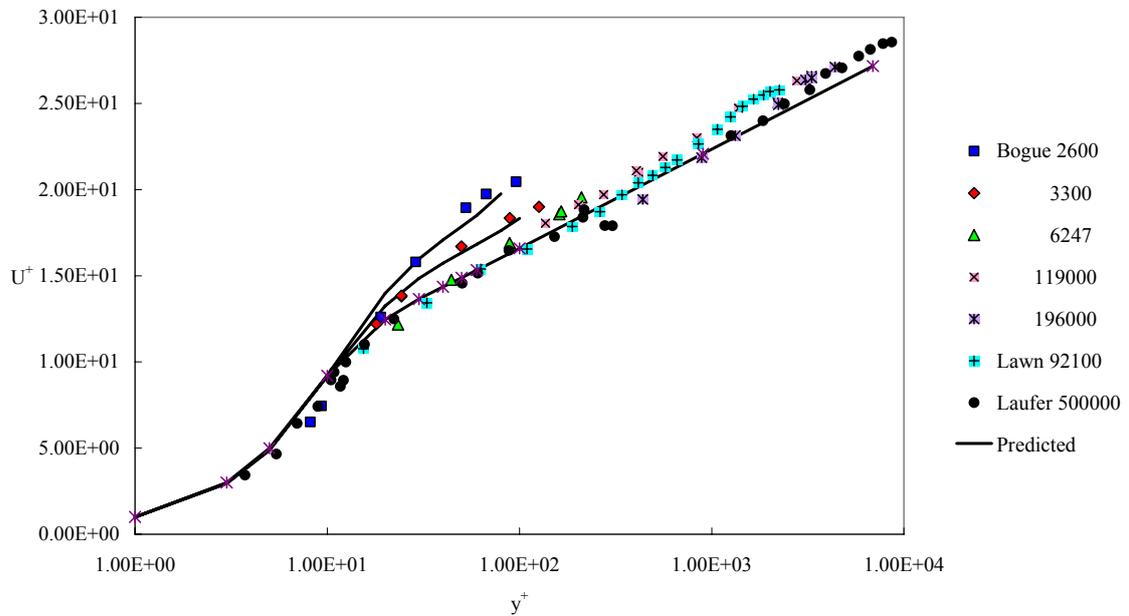

Figure 2. Predictions of velocity profiles from equation (40). Data of Bogue (op.cit.), Lawn (op.cit.) and Laufer (op.cit.)

The turbulent shear stress distribution is predicted by rearranging equation (29) as

$$\frac{\overline{U'V'}}{\tau} = \frac{\tau_t}{\tau} = 1 - \frac{dU^+/dy^+}{1 - y^+/R^+} \qquad (42)$$

Equation (42) is plotted against typical literature data in Figure 3.

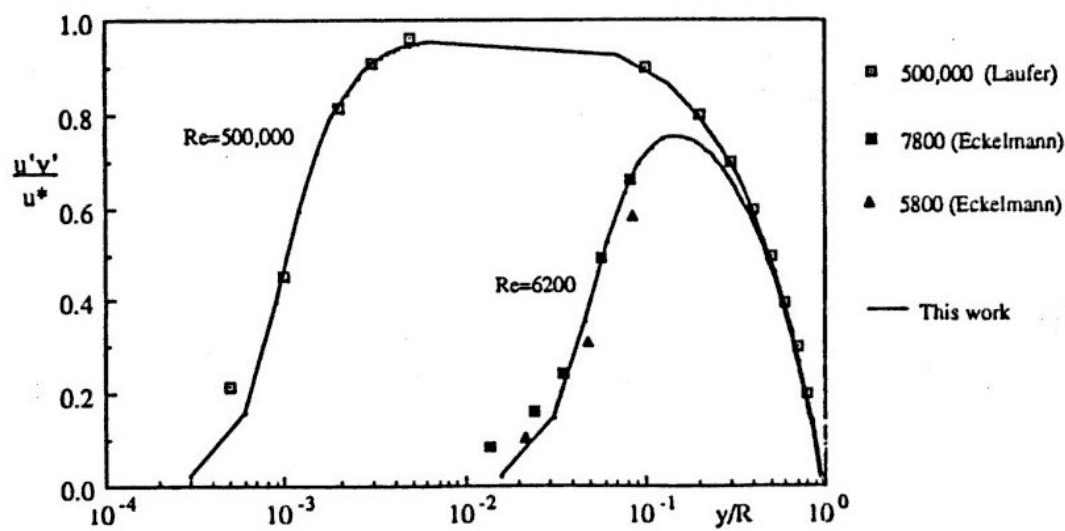

Figure 3. Reynolds stress distribution predicted by equation (42). Data of Laufer (op.cit.) and Eckelmann (1974)

The normalised turbulence dissipation is defined as

$$\varepsilon = \left[\frac{\nu}{\tau_w}\frac{dU}{dy}\right]^2 \qquad (43)$$

Which can be rearranged as

$$\varepsilon = \left[\frac{1 - y^+/R^+}{(E_\nu/\nu) + 1}\right] \qquad (44)$$

Equation (44) is plotted in Figure 4.

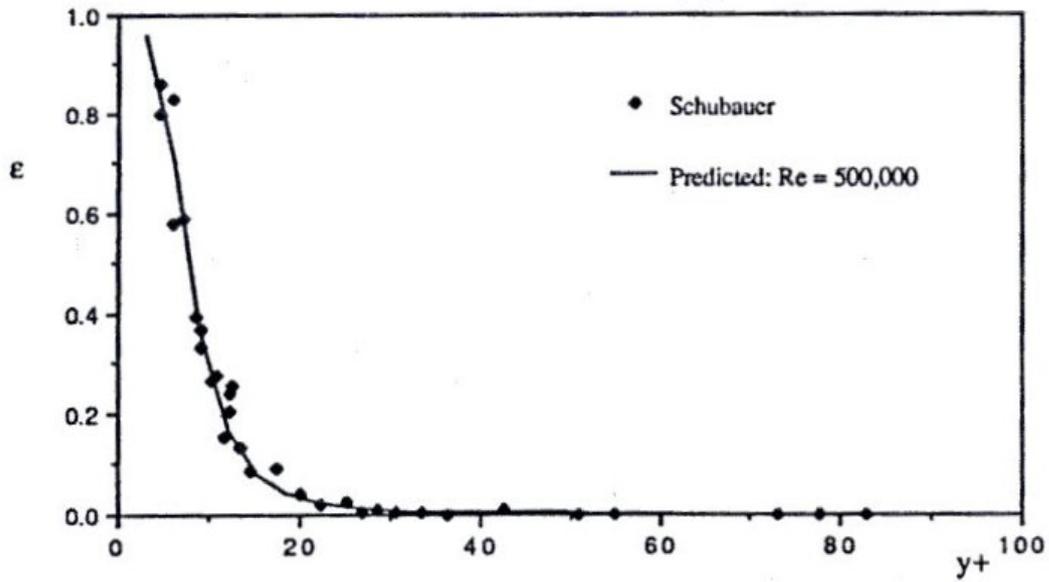

Figure 4. Dissipation of turbulence from equation (44). Data of Schubauer (1954)

The production of turbulent energy is defined as

$$P = \overline{U'V'}\frac{dU}{dy} \qquad (45)$$

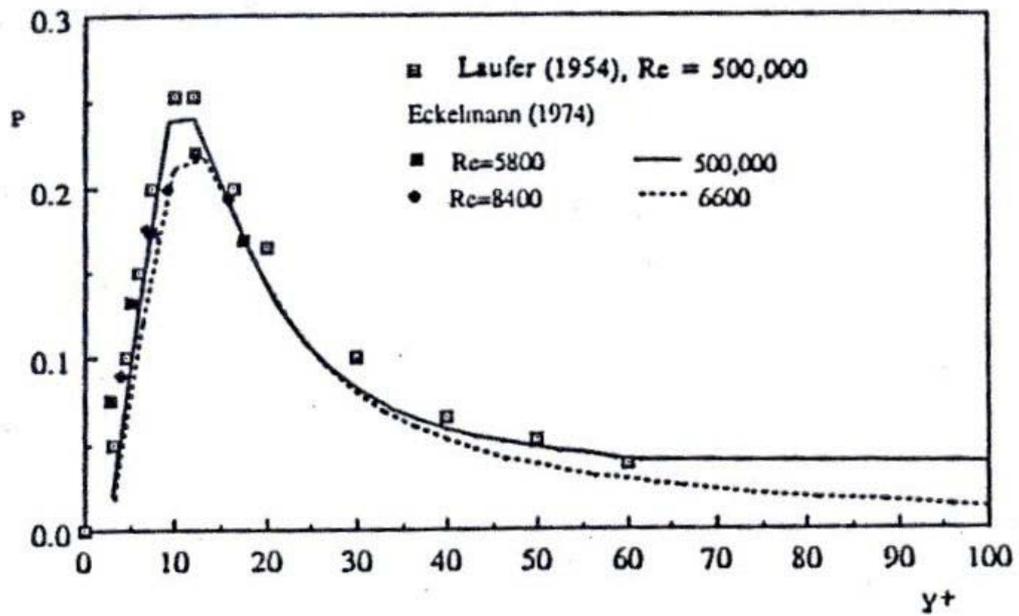

Figure 5. Production of turbulence predicted by equation (46). Data of Eckelmann (op.cit.) and Laufer (op.cit.)

$$\frac{P}{\nu} = \left(1 - \frac{y^+}{R^+}\right)\frac{dU^+}{dy^+} - \left(\frac{dU^+}{dy^+}\right)^2 \tag{46}$$

It is plotted in Figure 5.

## 5    Discussion

There are many other possible forms of equation (36) where the original formulation of the length scale was modified with a damping function to give a formulation that is compatible with the solution of order $\varepsilon_0$ near the wall. The ratio $\sqrt{\tau/\tau_w}$ was introduced simply for convenience of mathematical computation. Dimensionally speaking, any other ratio of stresses could have been chosen. For example we may have considered

$$l = \kappa y F \sqrt{\frac{\tau}{\tau_t}} \tag{47}$$

which leads to

$$\frac{E_\nu}{\nu} = \kappa y^+ F \sqrt{\frac{\tau}{\tau_w}} \tag{48}$$

Or

$$l = \kappa y F \sqrt{\frac{\tau_w}{\tau_t}} \tag{49}$$

$$\frac{E_\nu}{\nu} = \kappa y^+ F \tag{50}$$

Or

$$l = \kappa y F \tag{51}$$

$$\frac{E_\nu}{\nu} = \kappa y^+ F \sqrt{\frac{\tau_t}{\tau_w}} \tag{52}$$

With each of these formulations, the constant $b$ in equation (40) needs to be modified for best fit against the experimental data then all formulations give the same type of prediction shown in Figure 2. Real differences in the predictions of the eddy viscosity only become apparent when $y^+ \leq 0.1$ (Trinh, unpublished work, 1986; 1991) but these have negligible impact on the time averaged velocity profile or the friction factor predicted.

The log law is based on using only the first term in the Taylor series in equation (27). It will not apply when the second and higher order terms cannot be neglected. Therefore a matching criterion for the log-law layer and the outer region can be derived by writing

$$l\frac{dU}{dy} = l^2 \frac{d^2U}{dy^2} \tag{53}$$

or

$$l = \frac{dU/dy}{d^2U/dy^2} \tag{54}$$

Equation (54) can be recognised as Karman's similarity hypothesis. Matching equations (36) with $F = 1$ and (54) gives a formal estimate of the interface between the inner and outer regions

$$\kappa y \sqrt{\frac{\tau}{\tau_w}} = \frac{dU/dy}{d^2U/dy^2} \tag{55}$$

In older standard $k - \varepsilon$ models the switch from the iteration model to the logarithmic law of the wall is performed quite arbitrarily. The outer limit of the log-law varies considerably with the geometry and conditions of the outer flow (Trinh, 2010e, Trinh, 2010a). In some transitional flows, it may not even exist and the outer flow model should be coupled directly to the wall layer (Trinh, 2010e, Trinh, 2010d). Therefore a fixed arbitrary interface will add significant uncertainties to the solution obtained. In the outer region the effect of viscosity is very small and the flow is close to potential. In such situation the pressure term dominates and, as pointed by Bradshaw et al. (1991), most models perform quite well. The great problem for these models comes from formulating the Reynolds stresses closer to the wall. The use of equation (55) to switch to the log-law sidesteps that problem.

The parameters of the log-law for any particular situation can be determined accurately by forcing it through the edge of the wall layer, equation (21). The solution of order $\varepsilon^0$ is exact since it does not require any modelling of the Reynolds stresses. Therefore this match provides the missing degree of freedom into the RANS solution that is the log-law. The Lagrangian solution following the path of diffusion can be transformed into the Eulerian framework using the procedure in Trinh (2010b). It provides an estimate of the wall shear stress, equation (18), and then of the term $R^+$ for closed conduits or $\delta^+$ for external boundary layer flow to begin the next iteration. Because the iteration in the wall layer is based on an analytical solution like equations (17) and (18) it is much less demanding in computer time than iterations based on the original RANS.

## 6    Conclusion

The matching of two different descriptions of the wall layer, one unsteady state and one steady state leads to a successful closure of turbulence models in the inner region of turbulent flow fields. An example has been given for pipe flow. In this paper, the simplest model for an unsteady viscous sub-boundary layer at the wall, the Stokes solution, has been used successfully. However, the closure principle proposed here can be used with any other model for an unsteady state viscous layer.

The technique is based on a zonal similarity analysis of turbulent flow fields, which identifies different regions that are populated by different types of coherent structures. The method captures information about these regions, in particular their scales, from the unsteady state solutions and feed them back into the steady state models. The interfaces between these regions are located much more formally than before. In order to capture the effect of the Reynolds number, it is important to avoid some of the gross, and unnecessary, simplifications made, for example in the original derivation of the log-law by Prandtl.